\documentclass[pdflatex,sn-mathphys]{sn-jnl}

\jyear{2022}%

\theoremstyle{thmstyleone}%
\theoremstyle{thmstyletwo}%

\theoremstyle{thmstylethree}

\newcommand{\ith}{\ensuremath{^{\rm th}}}

\newcommand{\feh}{[Fe/H]}

\newcommand{\prot}{$P_{\rm rot}$}
\newcommand{\teff}{$T_{\rm eff}$}
\newcommand{\bprp}{$G_{\rm BP}-G_{\rm RP}$}

\usepackage{lineno}

\begin{document}

\title[Abrupt change in stellar spin-down law]{An abrupt change in the stellar spin-down law at the fully convective boundary}


\author*[1,2]{\sur{Lucy(Yuxi) Lu}}\email{yl4331@columbia.edu}
\author[3]{\sur{Victor See}}\email{victor.see@esa.int}
\author[4]{\sur{Louis Amard}}\email{louis.amard@cea.fr}
\author[1,2,5]{\sur{Ruth Angus}}\email{rangus@flatironinstitute.org}
\author[6]{\sur{Sean P. Matt}}\email{S.Matt@exeter.ac.uk}


\affil[1]{\orgdiv{Astronomy Department}, \orgname{Columbia University}, \orgaddress{\street{550 West 120\ith\ Street}, \city{New York}, \postcode{10027}, \state{New York}, \country{USA}}}

\affil[2]{\orgdiv{Astrophysics Department}, \orgname{American Museum of Natural History}, \orgaddress{\street{200 Central Park West}, \city{New York}, \postcode{10024}, \state{NY}, \country{USA}}}

\affil[3]{\orgname{European Space Agency (ESA), European Space Research and Technology Centre (ESTEC)}, \orgaddress{Keplerlaan 1}, \city{Noordwijk}, \postcode{2201 AZ}, \country{The Netherlands}}

\affil[4]{\orgname{AIM, CEA, CNRS}, \orgaddress{Universit\'e Paris-Saclay, Universit\'e de Paris, Sorbonne Paris Cit\'e}, \city{Gif-sur-Yvette}, \postcode{91191}, \country{France}}

\affil[5]{\orgdiv{Center for Computational Astrophysics}, \orgname{Flatiron Institute}, \orgaddress{\street{162 5th Avenue}, \city{New York}, \postcode{10010}, \state{NY}, \country{USA}}}

\affil[6]{\orgdiv{Department of Physics and Astronomy}, \orgname{University of Exeter}, \orgaddress{Physics Building, Stocker Road}, \city{Exeter}, \postcode{EX4 4QL}, \country{UK}}

\maketitle
{\bf The importance of the existence of a radiative core in generating a solar-like magnetic dynamo is still unclear. Analytic models and magnetohydrodynamic simulations of stars suggest the thin layer between a star's radiative core and its convective zone can produce shearing that reproduces key characteristics of a solar-like dynamo.   However, recent studies suggest fully and partially convective stars exhibit very similar period-activity relations, hinting that dynamos generated by stars with and without radiative cores hold similar properties. Here, using kinematic ages, we discover an abrupt change in the stellar spin-down law across the fully convective boundary. We found that fully convective stars exhibit a higher angular momentum loss rate, corresponding to a torque that is $\sim$ 2.25 times higher for a given angular velocity than partially convective stars around the fully convective boundary. This requires a dipole field strength that is larger by a factor of $\sim$2.5, a mass loss rate that is $\sim$4.2 times larger, or some combination of both of those factors. Since stellar-wind torques depend primarily on large-scale magnetic fields and mass loss rates, both of which derive from magnetic activity, the observed abrupt change in spin-down law suggests that the dynamos of partially and fully convective stars may be fundamentally different.}

The interiors of fully convective stars \cite[$M < \sim0.35M_\odot$;][]{Chabrier1997} and partially convective stars such as the Sun are fundamentally different, as fully convective stars do not possess a radiative core.
How the stellar magnetic dynamo is affected by the existence of a radiative core and more importantly, the role of the tachocline (the transition region between a star's radiative core and its convective zone) is still unclear. 
Models suggest that magnetic fields in sun-like stars are amplified and generated at the tachocline, which is defined by a shear layer between the core and the convective envelope \cite[e.g.][]{Spiegel1972,SpiegelZahnZ1992,DikpatiCharbonneau1999,Miesch2005}.
This implies dynamos in fully convective stars, which do not possess such shear layers, should be different \cite[e.g.][]{Durney1993,Bice2020}. 
It is also well known that the rotation periods and magnetic activities of stars are tightly correlated \citep[\it{e.g.}][]{Pallavicini1981}, and recent observational studies reveal fully convective stars exhibit similar period-activity relations to stars with radiative cores \cite[e.g.][]{Wright2016, Stelzer2016, Newton2017, Wright2018}, hinting that a radiative core may not be a critical ingredient for generating a sun-like dynamo. 

One way to break this tension between theoretical predictions and observations is by understanding the time evolution of rotation periods of stars on either side of the fully convective boundary.
Since a dynamo is ultimately responsible for generating the surface magnetic fields and the magnetic activity that drives stellar winds, the amount of angular momentum carried away by stellar winds should be sensitive to the details of dynamo processes.
The evolution of rotation rate is sensitive to long-timescale trends in the average wind torque, which thus probes trends in the global magnetic field strength and mass loss rate, both of which should be tied to global dynamo relationships.
Thus, understanding the spin-down law across the fully convective boundary could be the key to revealing the magnetic properties of stars and resolving this discrepancy.

Theoretical rotation evolution models that are constrained by observed rotation period distributions have provided insight into the magnetic topology and angular momentum transport in stars \cite[e.g.][]{Kraft1967, Kawaler1988, Krishnamurthi1997, matt2015, Amard2016, vansaders2016, Garraffo2018, Spada2020}.
However, most of these works have focused on understanding FGK dwarfs, as both periods and ages for old M dwarfs are extremely difficult to obtain.
Old M dwarfs are faint and many rotate slowly ($>$ 25 days). 
This means photometric data with high sensitivity and a long observational baseline are in need to measure their periods. 
Ages for old M dwarfs are also hard to infer as their observables change slowly with time, creating challenges to age-date them with isochrone fitting. 

Various studies have provided hints on the spin-down of these low-mass M dwarfs toward older ages. Galactic kinematic and wide binaries on a relatively small sample of M dwarfs with periods obtained from MEarth \citep{Irwin2011, Berta2012, Newton2018, Pass2022} found a bi-modality of fast and slow-rotating M dwarfs that is difficult to explain with traditional models of angular-momentum loss.
However, the rotation periods measured for the 4 Gyr open cluster M67 \citep{Dungee2022} suggest old M dwarfs do eventually converge onto a tight sequence.
However, we still lack the sample size, especially at older ages, to constrain a spin-down law of these fully convective stars from observational data that can be used to test theoretical models. 

A recent catalog of rotation periods measured using the Zwicky Transient Facility (ZTF) has the sample size needed to understand the spin-down of fully convective stars \citep{Lu2022}. 
With the gyro-kinematic age-dating method \citep{Angus2020, Lu2021}, we obtain kinematic ages for Kepler and ZTF stars with period measurements (See section~\ref{subsection:method} for more details).
We extend the age measurements for fully convective stars up to $\sim$ 10 Gyr and detect an abrupt change in the spin-down law across the fully convective boundary. 
\\
\\

{\bf\large Results}\\

\begin{figure}[h!]
    \centering
    \includegraphics[width=0.48\textwidth]{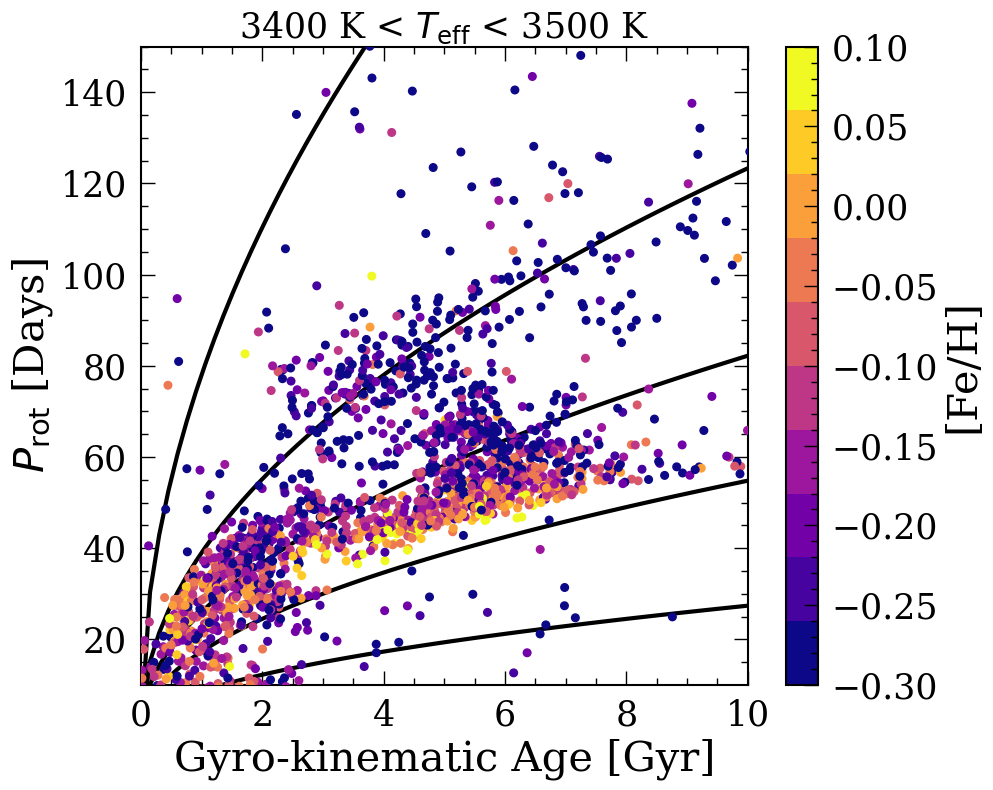}
    \includegraphics[width=0.48\textwidth]{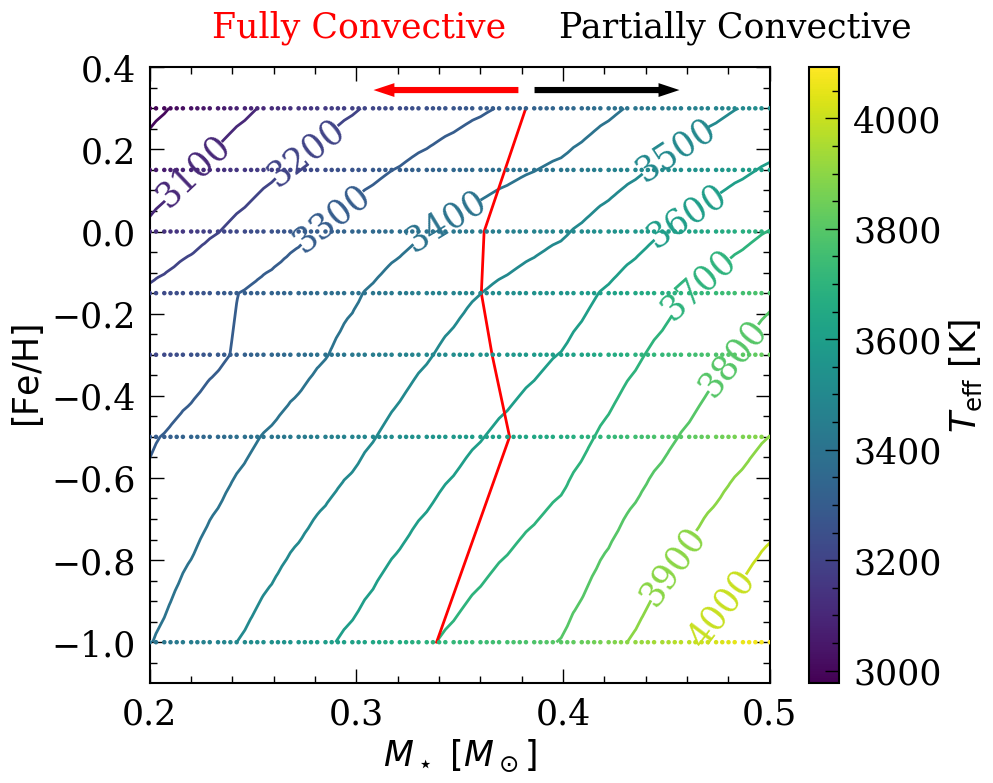}
    \caption{
    Left: observational data showing the \prot-age relation for stars between 3400 K $<$ \teff\ $<$ 3500 K colored by \feh.
    Right: The fully convective boundary predicted by the STAREVOL \citep{Siess2000,Amard2019} stellar evolution model for stars with various \teff\ and \feh. 
    The model suggests stars with the same \teff\ but lower \feh\ are fully convective while those with higher \feh\ are partially convective. 
    Both partially convective and fully convective stars should exist between $\sim$ 3400 K - 3600 K.
    The bi-modal distribution of rotation periods across stars of different metallicities seen in the left panel suggests the angular momentum loss rate changes abruptly across the fully convective boundary, in which the fully convective stars (top sequence; lower \feh) have a higher angular momentum loss rate compared to the partially convective stars (bottom sequence; higher \feh).}
    \label{fig:1}
\end{figure}

\textbf{A Double Sequence in a Narrow-\teff\ Bin Near the Fully Convective Boundary.}
The left panel of Fig.\ref{fig:1} shows the \prot-age relation for stars between 3400 K - 3500 K, colored by metallicity \citep{Andrae2023}.
A bi-modality emerges in the \prot-age relation in this temperature range with metal-poor stars spinning down quicker than metal-rich stars.

Stars with similar temperatures can have either fully or partially convective interiors depending on their metallicity.
Fig.~\ref{fig:1} (right plot) shows the fully convective boundary (red line) for stars with different \feh\ and temperatures predicted from the STAREVOL stellar evolution model \citep[][for details, see section~\ref{subsection:model}]{Siess2000, Amard2019}.
The model suggests that, near the fully convective boundary, stars with the same temperature but lower \feh\ are fully convective while those with higher \feh\ remain partially convective, and that both fully and partially convective stars exist between $\sim$ 3400 K - 3600 K.
This means, given a sample of stars with a wide range of \feh\footnote{Since the ZTF survey covers the entire northern sky, this statement should be satisfied.}, we can compare the \prot-age relation for both fully and partially convective stars with similar masses simultaneously by selecting stars in a narrow range of temperatures near the fully convective boundary.
Since the upper sequence in Fig.~\ref{fig:1} left plot is more metal-poor than the lower sequence, we interpret that the upper sequence consists of FC stars and the lower sequence consists of PC stars.


\textbf{The Bi-modality of Spin-down Laws for Fully/Partially Convective Stars.}
To further understand whether this bi-modality is actually caused by an abrupt change in the spin-down law of stars across the fully convective boundary, we separate the fully and partially convective stars based on their absolute Gaia magnitude, $M_G$, and Gaia BP-RP color measurements using Jao's gap \citep{Jao2018}. 
We use the Jao gap to separate the stars as the measurement for $M_G$ and Gaia BP-RP color are reliable.
Jao's gap is an under-density in the color-magnitude-diagram (CMD) near the fully convective boundary discovered using stars within 200 pc of the Sun from Gaia DR2 \citep{gaia, gaia2018}.
This gap can be approximated by a line connecting ($M_G$, \bprp) $\sim$ (10.09 mag, 2.35 mag) and ($M_G$, \bprp) $\sim$ (10.24 mag, 2.55 mag), and is thought to be caused by structural instabilities due to non-equilibrium fusion of $^3$He \citep{vansanders2012, Baraffe2018, MacDonald2018, Feiden2021}.
In the rest of the paper, we use this line to roughly separate the fully and partially convective stars, in which stars lying above this line in the CMD are most likely partially convective, and those below are most likely fully convective. 

\begin{figure*}
    \centering
    \includegraphics[width=\textwidth]{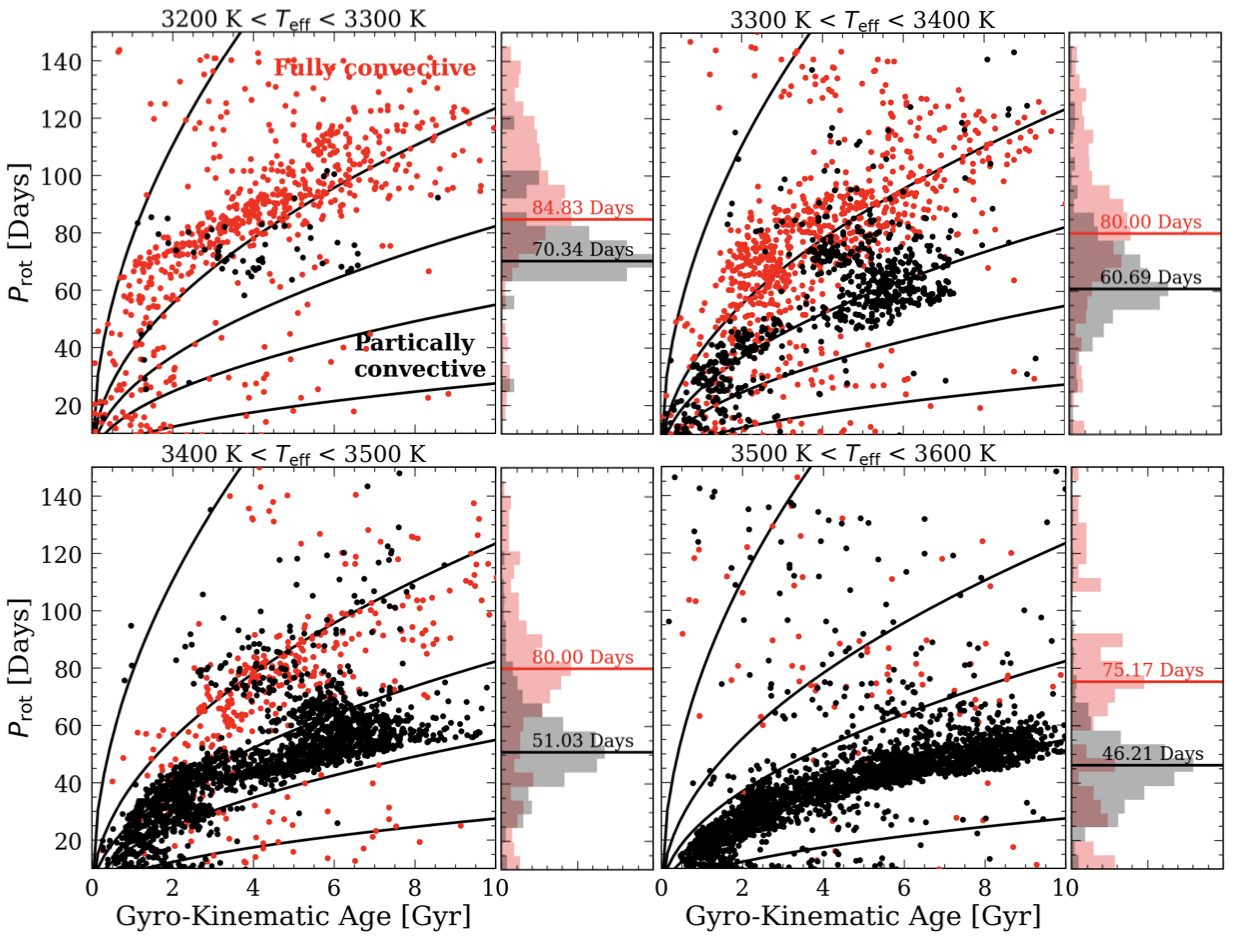}
    \caption{\prot-age relation for stars in 4 different narrow \teff\;bins, where the fully convective stars (red) and the partially convective stars (black) are separated using the Jao's gap \citep{Jao2018}.
    The four black lines are Skumanich spin-down laws \citep[\prot $\propto$ Age$^{0.5}$;][]{Skumanich1972} to guide the eyes, these lines show that, at a given age, the fully convective stars have a spin period that is $\sim$1.5 times larger than the partially convective stars.
    The normalized histograms on the right of each subplot are the period distributions for fully convective stars (red) and partially convective stars (black) that has gyro-kinematic ages $>$ 2 Gyr.}
    \label{fig:2}
\end{figure*}

We plot the \prot-age relation for fully (red) and partially (black) convective stars in narrow temperature bins of 100 K, between 3200 K to 3600 K (Fig.\ref{fig:2}). 
The normalized histograms on the right show the bi-modal period distributions of these stars with gyro-kinematic age $>$ 2 Gyr, and the lines mark the bins with the highest normalized number density.
It is notable that the double sequence mostly exists for stars between 3300 K $<$ \teff\ $<$ 3500 K: this is slightly lower than, yet very close to, the \teff\ range that stellar evolution model predicts to contain both fully and partially convective stars if a sample contains stars with a wide range of \feh\ (Fig.\ref{fig:1}, right plot).
More interestingly, stars below 3300 K (fully convective) follow the top sequence, and those above 3500 K (partially convective) follow the bottom sequence, further suggesting the spin-down laws of fully and partially convective stars are bi-modal. 
However, it is worth pointing out that this gap does not provide a clean division between partially and fully convective stars as they can oscillate between this gap while transitioning between being partially and fully convective \citep{Baraffe2018}. 

The abrupt change in the rotational evolution between fully and partially convective stars means the angular momentum loss rates also exhibit an abrupt change between stars with and without a radiative core. 
As shown by the detailed calculation in section.~\ref{subsection:calculation}, a fully convective star experiences a higher spin-down torque by a factor of $\sim$2.25 than a partially convective star with the same rotation period.  Stellar wind theory then predicts that, at a given rotation period, fully convective stars should have dipole fields that are $\sim$2.5 times stronger or mass-loss rates that are $\sim$4.2 times higher (or some combination of both).

\begin{figure*}
    \centering
    \includegraphics[width=\textwidth]{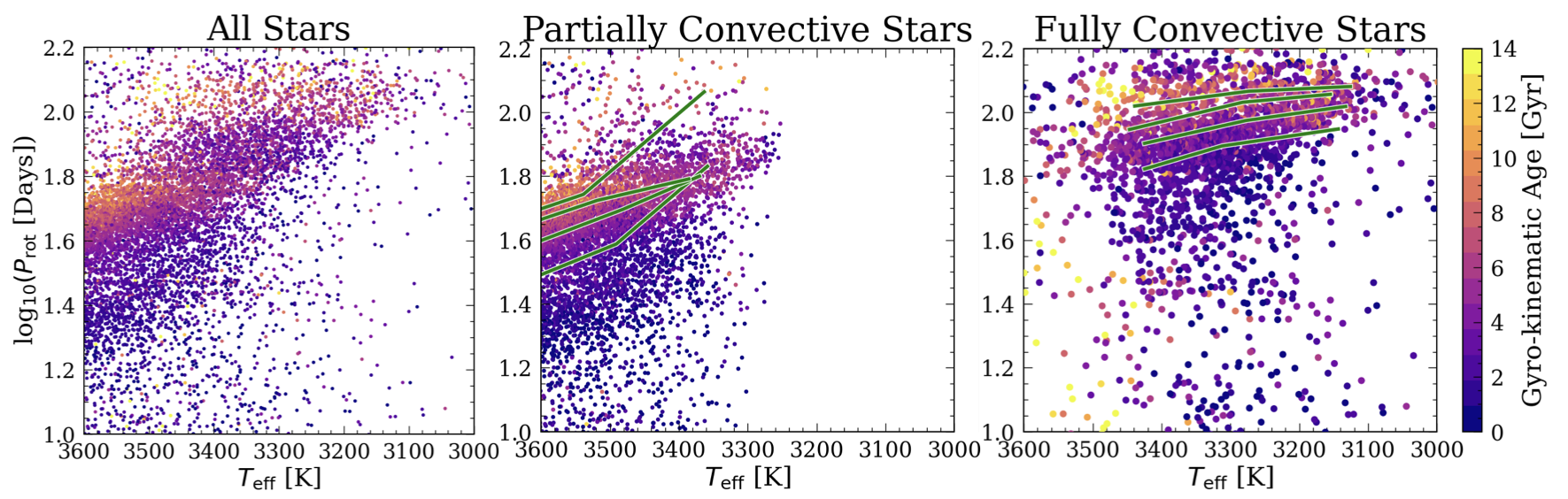}
    \caption{log$_{10} (P_{\rm rot})$-\teff\ distribution for the entire sample (left), partially convective stars (middle), and fully convective stars (right) colored by gyro-kinematic ages.
    The green solid lines are the running median for stars in mono-age bins of 2 Gyr (equivalent to isochrones). 
    Fully convective stars are younger compared to partially convective stars at the same period range, suggesting fully convective stars have a higher angular momentum loss rate.
    Flatter isochrones for the fully convective stars compared to those for the partially convective stars suggest the angular momentum loss rate is less \teff-dependent for the former.}
    \label{fig:3}
\end{figure*}

\textbf{The Rotation Period Distributions of Fully/Partially Convective Stars.}
With gyro-kinematic ages, we can directly examine the rotation period distributions and spin-evolution isochrones for partially/fully convective stars. 
Fig.\ref{fig:3} shows the log$_{10}(P_{\rm rot})$-\teff\ diagram for all the stars (left), partially convective stars (middle), and fully convective stars (right) colored by gyro-kinematic ages. 
The age distribution for the entire sample suggests for a fixed temperature, stars that spin slower are normally older.
However, around the fully convective boundary, this trend no longer holds. 
This discrepancy exists because partially and fully convective stars belong to different spin-down sequences.
Within their own slow-rotating sequence, slow rotators are indeed older than fast rotators.
Moreover, the slopes of the isochrones (Solid green lines in the middle and the right plot of Fig.\ref{fig:3}) for fully convective stars are more shallow compared to those for the partially convective stars, indicating the spin-down law for the former is less \teff-dependent, indicating a weaker correlation between \teff\ and angular momentum loss rate for fully convective stars compared to that for partially convective stars. 
The intermediate period gap, an observed dearth of stellar rotation periods in the temperature--period diagram at $\sim$ 20 days for G dwarfs and up to $\sim$ 30 days for early-M dwarfs, only appears in the partially convective stars, further supporting the hypothesis that this period gap is formed from the re-distribution of angular momentum between a star's radiative core and its convective envelope \citep{Curtis2020, Spada2020, Gordon2021, Lu2022}.

{\bf\large Discussion}\\
The bi-modality of spin-down laws for fully and partially convective stars means they have fundamentally different spin-down laws and thus, different angular momentum loss rates.
Observational data suggests fully convective stars lose angular momentum $\sim$2.25 times faster than partially convective stars at a given angular velocity (see section~\ref{subsection:calculation} for more detail on this calculation). 
Since stellar spin-down and winds are direct consequences of the stellar dynamo, this suggests the dynamos of fully convective and partially convective stars are also fundamentally different.
This result is the most pronounced observational evidence to date that the dynamos of fully convective and partially convective stars operate differently.
However, the exact operational difference between their dynamos is not clear as dynamos and the mechanisms that convert magnetic energy into the heating that drives stellar winds and angular momentum loss are still not well understood.

Typically, stellar-wind theory indicates that the angular momentum loss rate should be directly correlated with the mass-loss rates, the strength, and the geometry of the magnetic fields \citep[e.g.][]{Mestel1984, Kawaler1988}.
Observations \cite{Wood2021} suggest the mass-loss rate for fully convective M dwarfs is similar or smaller than that of the Sun based on the UV spectra of stellar HI Ly$\alpha$ lines from the Hubble Space Telescope. 
Activity indicators such as X-ray \citep[e.g.][]{Wright2016, Wright2018} and H$\alpha$ \citep[e.g.][]{Newton2017, Anthony2022} also suggest fully and partially convective stars exhibit similar $Ro$-activity relations, indicating M dwarfs generate solar-like dynamos. 
These results contradict the drastic change of the rotation evolution across the fully convective boundary discovered in this paper.
However, since the scatter around the $Ro$-activity relation is high, further observations to reduce the uncertainty of these $Ro$-activity are needed to draw a definite conclusion. 

To reconcile these results, it is important to understand that most magnetic indicators (e.g. X-ray, H$\alpha$) probe active regions of a star that are mostly generated by small-scale magnetic fields. 
However, angular momentum loss through magnetic winds (stellar spin-down) is likely driven by escaping of open field lines produced by the magnetic dipole (large-scale magnetic field) \citep{See2019,See2020}. 
As a result, combining prior studies and this work, we speculate the differences in the dynamos of fully convective and partially convective stars exist in their magnetic morphology, which causes fully convective stars to generate magnetic dipoles that are stronger but similar higher-order magnetic fields compared to partially convective stars. 

\backmatter

\bmhead{Acknowledgments}
Y.L. acknowledges support from ESA through the Science Faculty of the European Space Research and Technology Centre (ESTEC). V.S. acknowledges support from the European Space Agency (ESA) as an ESA Research Fellow.  R.A. acknowledges support from NSF AAG grant \#2108251. S.P.M. acknowledges support as a visiting scholar from the Center for Computational Astrophysics at the Flatiron Institute, which is supported by the Simons Foundation.

This work has made use of data from the European Space Agency (ESA)
mission Gaia,\footnote{\url{https://www.cosmos.esa.int/gaia}} processed by
the Gaia Data Processing and Analysis Consortium (DPAC).\footnote{\url{https://www.cosmos.esa.int/web/gaia/dpac/consortium}} Funding
for the DPAC has been provided by national institutions, in particular
the institutions participating in the Gaia Multilateral Agreement.
This research also made use of public auxiliary data provided by ESA/Gaia/DPAC/CU5 and prepared by Carine Babusiaux.

This research was done using services provided by the OSG Consortium \citep{OSG1,OSG2}, which is supported by the National Science Foundation awards \#2030508 and \#1836650.

This research has also made use of NASA's Astrophysics Data System, 
and the VizieR \citep{vizier} and SIMBAD \citep{simbad} databases, 
operated at CDS, Strasbourg, France.

\bibliography{sn-article}


\begin{thebibliography}{62}
\ifx \bisbn   \undefined \def \bisbn  #1{ISBN #1}\fi
\ifx \binits  \undefined \def \binits#1{#1}\fi
\ifx \bauthor  \undefined \def \bauthor#1{#1}\fi
\ifx \batitle  \undefined \def \batitle#1{#1}\fi
\ifx \bjtitle  \undefined \def \bjtitle#1{#1}\fi
\ifx \bvolume  \undefined \def \bvolume#1{\textbf{#1}}\fi
\ifx \byear  \undefined \def \byear#1{#1}\fi
\ifx \bissue  \undefined \def \bissue#1{#1}\fi
\ifx \bfpage  \undefined \def \bfpage#1{#1}\fi
\ifx \blpage  \undefined \def \blpage #1{#1}\fi
\ifx \burl  \undefined \def \burl#1{\textsf{#1}}\fi
\ifx \doiurl  \undefined \def \doiurl#1{\url{https://doi.org/#1}}\fi
\ifx \betal  \undefined \def \betal{\textit{et al.}}\fi
\ifx \binstitute  \undefined \def \binstitute#1{#1}\fi
\ifx \binstitutionaled  \undefined \def \binstitutionaled#1{#1}\fi
\ifx \bctitle  \undefined \def \bctitle#1{#1}\fi
\ifx \beditor  \undefined \def \beditor#1{#1}\fi
\ifx \bpublisher  \undefined \def \bpublisher#1{#1}\fi
\ifx \bbtitle  \undefined \def \bbtitle#1{#1}\fi
\ifx \bedition  \undefined \def \bedition#1{#1}\fi
\ifx \bseriesno  \undefined \def \bseriesno#1{#1}\fi
\ifx \blocation  \undefined \def \blocation#1{#1}\fi
\ifx \bsertitle  \undefined \def \bsertitle#1{#1}\fi
\ifx \bsnm \undefined \def \bsnm#1{#1}\fi
\ifx \bsuffix \undefined \def \bsuffix#1{#1}\fi
\ifx \bparticle \undefined \def \bparticle#1{#1}\fi
\ifx \barticle \undefined \def \barticle#1{#1}\fi
\bibcommenthead
\ifx \bconfdate \undefined \def \bconfdate #1{#1}\fi
\ifx \botherref \undefined \def \botherref #1{#1}\fi
\ifx \url \undefined \def \url#1{\textsf{#1}}\fi
\ifx \bchapter \undefined \def \bchapter#1{#1}\fi
\ifx \bbook \undefined \def \bbook#1{#1}\fi
\ifx \bcomment \undefined \def \bcomment#1{#1}\fi
\ifx \oauthor \undefined \def \oauthor#1{#1}\fi
\ifx \citeauthoryear \undefined \def \citeauthoryear#1{#1}\fi
\ifx \endbibitem  \undefined \def \endbibitem {}\fi
\ifx \bconflocation  \undefined \def \bconflocation#1{#1}\fi
\ifx \arxivurl  \undefined \def \arxivurl#1{\textsf{#1}}\fi
\csname PreBibitemsHook\endcsname

\bibitem{Chabrier1997}
\begin{barticle}
\bauthor{\bsnm{Chabrier}, \binits{G.}},
\bauthor{\bsnm{Baraffe}, \binits{I.}}:
\batitle{{Structure and evolution of low-mass stars}}.
\bjtitle{Astronomy and Astrophysics}
\bvolume{327},
\bfpage{1039}--\blpage{1053}
(\byear{1997})
\end{barticle}
\endbibitem

\bibitem{Spiegel1972}
\begin{bchapter}
\bauthor{\bsnm{Spiegel}, \binits{E.A.}}:
\bctitle{{A History of Solar Rotation}}.
In: \bbtitle{NASA Special Publication}
vol. \bseriesno{300},
p. \bfpage{61}
(\byear{1972})
\end{bchapter}
\endbibitem

\bibitem{SpiegelZahnZ1992}
\begin{barticle}
\bauthor{\bsnm{{Spiegel}}, \binits{E.A.}},
\bauthor{\bsnm{{Zahn}}, \binits{J.-P.}}:
\batitle{{The solar tachocline.}}
\bjtitle{Astronomy and Astrophysics}
\bvolume{265},
\bfpage{106}--\blpage{114}
(\byear{1992})
\end{barticle}
\endbibitem

\bibitem{DikpatiCharbonneau1999}
\begin{barticle}
\bauthor{\bsnm{Dikpati}, \binits{M.}},
\bauthor{\bsnm{Charbonneau}, \binits{P.}}:
\batitle{{A Babcock-Leighton Flux Transport Dynamo with Solar-like Differential
  Rotation}}.
\bjtitle{Astrophysical Journal}
\bvolume{518}(\bissue{1}),
\bfpage{508}--\blpage{520}
(\byear{1999}).
\doiurl{10.1086/307269}
\end{barticle}
\endbibitem

\bibitem{Miesch2005}
\begin{barticle}
\bauthor{\bsnm{Miesch}, \binits{M.S.}}:
\batitle{{Large-Scale Dynamics of the Convection Zone and Tachocline}}.
\bjtitle{Living Reviews in Solar Physics}
\bvolume{2}(\bissue{1}),
\bfpage{1}
(\byear{2005}).
\doiurl{10.12942/lrsp-2005-1}
\end{barticle}
\endbibitem

\bibitem{Durney1993}
\begin{barticle}
\bauthor{\bsnm{{Durney}}, \binits{B.R.}},
\bauthor{\bsnm{{De Young}}, \binits{D.S.}},
\bauthor{\bsnm{{Roxburgh}}, \binits{I.W.}}:
\batitle{{On the Generation of the Largescale and Turbulent Magnetic Fields in
  the Solar Type Stars}}.
\bjtitle{Solar Physics}
\bvolume{145}(\bissue{2}),
\bfpage{207}--\blpage{225}
(\byear{1993}).
\doiurl{10.1007/BF00690652}
\end{barticle}
\endbibitem

\bibitem{Bice2020}
\begin{barticle}
\bauthor{\bsnm{Bice}, \binits{C.P.}},
\bauthor{\bsnm{Toomre}, \binits{J.}}:
\batitle{{Probing the Influence of a Tachocline in Simulated M-dwarf Dynamos}}.
\bjtitle{Astrophysical Journal}
\bvolume{893}(\bissue{2}),
\bfpage{107}
(\byear{2020}).
\doiurl{10.3847/1538-4357/ab8190}
\end{barticle}
\endbibitem

\bibitem{Pallavicini1981}
\begin{barticle}
\bauthor{\bsnm{Pallavicini}, \binits{R.}},
\bauthor{\bparticle{et} \bsnm{al.}}:
\batitle{{Relations among stellar X-ray emission observed from Einstein,
  stellar rotation and bolometric luminosity.}}
\bjtitle{Astrophysical Journal}
\bvolume{248},
\bfpage{279}--\blpage{290}
(\byear{1981}).
\doiurl{10.1086/159152}
\end{barticle}
\endbibitem

\bibitem{Wright2016}
\begin{barticle}
\bauthor{\bsnm{Wright}, \binits{N.J.}},
\bauthor{\bsnm{Drake}, \binits{J.J.}}:
\batitle{{Solar-type dynamo behaviour in fully convective stars without a
  tachocline}}.
\bjtitle{Nature}
\bvolume{535}(\bissue{7613}),
\bfpage{526}--\blpage{528}
(\byear{2016}).
\doiurl{10.1038/nature18638}
\end{barticle}
\endbibitem

\bibitem{Stelzer2016}
\begin{barticle}
\bauthor{\bsnm{Stelzer}, \binits{B.}},
\bauthor{\bsnm{Damasso}, \binits{M.}},
\bauthor{\bsnm{Scholz}, \binits{A.}},
\bauthor{\bsnm{Matt}, \binits{S.P.}}:
\batitle{{A path towards understanding the rotation-activity relation of M
  dwarfs with K2 mission, X-ray and UV data}}.
\bjtitle{\mnras}
\bvolume{463}(\bissue{2}),
\bfpage{1844}--\blpage{1864}
(\byear{2016})
\end{barticle}
\endbibitem

\bibitem{Newton2017}
\begin{barticle}
\bauthor{\bsnm{Newton}, \binits{E.R.}},
\bauthor{\bparticle{et} \bsnm{al.}}:
\batitle{{The H{\ensuremath{\alpha}} Emission of Nearby M Dwarfs and its
  Relation to Stellar Rotation}}.
\bjtitle{Astrophysical Journal}
\bvolume{834}(\bissue{1}),
\bfpage{85}
(\byear{2017}).
\doiurl{10.3847/1538-4357/834/1/85}
\end{barticle}
\endbibitem

\bibitem{Wright2018}
\begin{barticle}
\bauthor{\bsnm{Wright}, \binits{N.J.}},
\bauthor{\bparticle{et} \bsnm{al.}}:
\batitle{{The stellar rotation-activity relationship in fully convective M
  dwarfs}}.
\bjtitle{Monthly Notices of the Royal Astronomical Society}
\bvolume{479}(\bissue{2}),
\bfpage{2351}--\blpage{2360}
(\byear{2018}).
\doiurl{10.1093/mnras/sty1670}
\end{barticle}
\endbibitem

\bibitem{Kraft1967}
\begin{barticle}
\bauthor{\bsnm{Kraft}, \binits{R.P.}}:
\batitle{{Studies of Stellar Rotation. V. The Dependence of Rotation on Age
  among Solar-Type Stars}}.
\bjtitle{Astrophysical Journal}
\bvolume{150},
\bfpage{551}
(\byear{1967}).
\doiurl{10.1086/149359}
\end{barticle}
\endbibitem

\bibitem{Kawaler1988}
\begin{barticle}
\bauthor{\bsnm{{Kawaler}}, \binits{S.D.}}:
\batitle{{Angular Momentum Loss in Low-Mass Stars}}.
\bjtitle{Astrophysical Journal}
\bvolume{333},
\bfpage{236}
(\byear{1988}).
\doiurl{10.1086/166740}
\end{barticle}
\endbibitem

\bibitem{Krishnamurthi1997}
\begin{barticle}
\bauthor{\bsnm{Krishnamurthi}, \binits{A.}},
\bauthor{\bsnm{Pinsonneault}, \binits{M.H.}},
\bauthor{\bsnm{Barnes}, \binits{S.}},
\bauthor{\bsnm{Sofia}, \binits{S.}}:
\batitle{{Theoretical Models of the Angular Momentum Evolution of Solar-Type
  Stars}}.
\bjtitle{Astrophysical Journal}
\bvolume{480}(\bissue{1}),
\bfpage{303}--\blpage{323}
(\byear{1997})
\end{barticle}
\endbibitem

\bibitem{matt2015}
\begin{barticle}
\bauthor{\bsnm{Matt}, \binits{S.P.}},
\bauthor{\bparticle{et} \bsnm{al.}}:
\batitle{{The Mass-dependence of Angular Momentum Evolution in Sun-like
  Stars}}.
\bjtitle{Astrophysical Journall}
\bvolume{799}(\bissue{2}),
\bfpage{23}
(\byear{2015}).
\doiurl{10.1088/2041-8205/799/2/L23}
\end{barticle}
\endbibitem

\bibitem{Amard2016}
\begin{barticle}
\bauthor{\bsnm{Amard}, \binits{L.}},
\bauthor{\bparticle{et} \bsnm{al.}}:
\batitle{{Rotating models of young solar-type stars. Exploring braking laws and
  angular momentum transport processes}}.
\bjtitle{Astronomy and Astrophysics}
\bvolume{587},
\bfpage{105}
(\byear{2016}).
\doiurl{10.1051/0004-6361/201527349}
\end{barticle}
\endbibitem

\bibitem{vansaders2016}
\begin{barticle}
\bauthor{\bparticle{van} \bsnm{Saders}, \binits{J.L.}},
\bauthor{\bparticle{et} \bsnm{al.}}:
\batitle{{Weakened magnetic braking as the origin of anomalously rapid rotation
  in old field stars}}.
\bjtitle{Nature}
\bvolume{529}(\bissue{7585}),
\bfpage{181}--\blpage{184}
(\byear{2016}).
\doiurl{10.1038/nature16168}
\end{barticle}
\endbibitem

\bibitem{Garraffo2018}
\begin{barticle}
\bauthor{\bsnm{Garraffo}, \binits{C.}},
\bauthor{\bparticle{et} \bsnm{al.}}:
\batitle{{The Revolution Revolution: Magnetic Morphology Driven Spin-down}}.
\bjtitle{Astrophysical Journal}
\bvolume{862}(\bissue{1}),
\bfpage{90}
(\byear{2018})
\end{barticle}
\endbibitem

\bibitem{Spada2020}
\begin{barticle}
\bauthor{\bsnm{Spada}, \binits{F.}},
\bauthor{\bsnm{Lanzafame}, \binits{A.C.}}:
\batitle{{Competing effect of wind braking and interior coupling in the
  rotational evolution of solar-like stars}}.
\bjtitle{Astronomy and Astrophysics}
\bvolume{636},
\bfpage{76}
(\byear{2020}).
\doiurl{10.1051/0004-6361/201936384}
\end{barticle}
\endbibitem

\bibitem{Irwin2011}
\begin{barticle}
\bauthor{\bsnm{Irwin}, \binits{J.}},
\bauthor{\bparticle{et} \bsnm{al.}}:
\batitle{{On the Angular Momentum Evolution of Fully Convective Stars: Rotation
  Periods for Field M-dwarfs from the MEarth Transit Survey}}.
\bjtitle{Astrophysical Journal}
\bvolume{727}(\bissue{1}),
\bfpage{56}
(\byear{2011}).
\doiurl{10.1088/0004-637X/727/1/56}
\end{barticle}
\endbibitem

\bibitem{Berta2012}
\begin{barticle}
\bauthor{\bsnm{Berta}, \binits{Z.K.}},
\bauthor{\bparticle{et} \bsnm{al.}}:
\batitle{{Transit Detection in the MEarth Survey of Nearby M Dwarfs: Bridging
  the Clean-first, Search-later Divide}}.
\bjtitle{Astronomical Journal}
\bvolume{144}(\bissue{5}),
\bfpage{145}
(\byear{2012}).
\doiurl{10.1088/0004-6256/144/5/145}
\end{barticle}
\endbibitem

\bibitem{Newton2018}
\begin{barticle}
\bauthor{\bsnm{{Newton}}, \binits{E.R.}},
\bauthor{\bsnm{{Mondrik}}, \binits{N.}},
\bauthor{\bsnm{{Irwin}}, \binits{J.}},
\bauthor{\bsnm{{Winters}}, \binits{J.G.}},
\bauthor{\bsnm{{Charbonneau}}, \binits{D.}}:
\batitle{{New Rotation Period Measurements for M Dwarfs in the Southern
  Hemisphere: An Abundance of Slowly Rotating, Fully Convective Stars}}.
\bjtitle{\aj}
\bvolume{156}(\bissue{5}),
\bfpage{217}
(\byear{2018})
\end{barticle}
\endbibitem

\bibitem{Pass2022}
\begin{barticle}
\bauthor{\bsnm{Pass}, \binits{E.K.}},
\bauthor{\bsnm{Charbonneau}, \binits{D.}},
\bauthor{\bsnm{Irwin}, \binits{J.M.}},
\bauthor{\bsnm{Winters}, \binits{J.G.}}:
\batitle{{Constraints on the Spindown of Fully Convective M Dwarfs Using Wide
  Field Binaries}}.
\bjtitle{Astrophysical Journal}
\bvolume{936}(\bissue{2}),
\bfpage{109}
(\byear{2022}).
\doiurl{10.3847/1538-4357/ac7da8}
\end{barticle}
\endbibitem

\bibitem{Dungee2022}
\begin{barticle}
\bauthor{\bsnm{Dungee}, \binits{R.}},
\bauthor{\bparticle{et} \bsnm{al.}}:
\batitle{{A 4 Gyr M-dwarf Gyrochrone from CFHT/MegaPrime Monitoring of the Open
  Cluster M67}}.
\bjtitle{Astrophysical Journal}
\bvolume{938}(\bissue{2}),
\bfpage{118}
(\byear{2022}).
\doiurl{10.3847/1538-4357/ac90be}
\end{barticle}
\endbibitem

\bibitem{Lu2022}
\begin{barticle}
\bauthor{\bsnm{Lu}, \binits{Y.}},
\bauthor{\bparticle{et} \bsnm{al.}}:
\batitle{{Bridging the Gap-The Disappearance of the Intermediate Period Gap for
  Fully Convective Stars, Uncovered by New ZTF Rotation Periods}}.
\bjtitle{Astronomical Journal}
\bvolume{164}(\bissue{6}),
\bfpage{251}
(\byear{2022}).
\doiurl{10.3847/1538-3881/ac9bee}
\end{barticle}
\endbibitem

\bibitem{Angus2020}
\begin{barticle}
\bauthor{\bsnm{Angus}, \binits{R.}},
\bauthor{\bparticle{et} \bsnm{al.}}:
\batitle{{Exploring the Evolution of Stellar Rotation Using Galactic
  Kinematics}}.
\bjtitle{Astronomical Journal}
\bvolume{160}(\bissue{2}),
\bfpage{90}
(\byear{2020}).
\doiurl{10.3847/1538-3881/ab91b2}
\end{barticle}
\endbibitem

\bibitem{Lu2021}
\begin{barticle}
\bauthor{\bsnm{Lu}, \binits{Y.}},
\bauthor{\bparticle{et} \bsnm{al.}}:
\batitle{{Gyro-kinematic Ages for around 30,000 Kepler Stars}}.
\bjtitle{Astronomical Journal}
\bvolume{161}(\bissue{4}),
\bfpage{189}
(\byear{2021}).
\doiurl{10.3847/1538-3881/abe4d6}
\end{barticle}
\endbibitem

\bibitem{Andrae2023}
\begin{botherref}
\oauthor{\bsnm{Andrae}, \binits{R.}},
\oauthor{\bsnm{Rix}, \binits{H.-W.}},
\oauthor{\bsnm{Chandra}, \binits{V.}}:
{Robust Data-driven Metallicities for 120 Million Stars from Gaia XP Spectra}.
arXiv e-prints,
2302--02611
(2023)
{\href{https://arxiv.org/abs/2302.02611}{{arXiv:2302.02611}}}
{[astro-ph.SR]}.
\doiurl{10.48550/arXiv.2302.02611}
\end{botherref}
\endbibitem

\bibitem{Siess2000}
\begin{barticle}
\bauthor{\bsnm{{Siess}}, \binits{L.}},
\bauthor{\bsnm{{Dufour}}, \binits{E.}},
\bauthor{\bsnm{{Forestini}}, \binits{M.}}:
\batitle{{An internet server for pre-main sequence tracks of low- and
  intermediate-mass stars}}.
\bjtitle{Astronomy and Astrophysics}
\bvolume{358},
\bfpage{593}--\blpage{599}
(\byear{2000})
\end{barticle}
\endbibitem

\bibitem{Amard2019}
\begin{barticle}
\bauthor{\bsnm{Amard}, \binits{L.}},
\bauthor{\bparticle{et} \bsnm{al.}}:
\batitle{{First grids of low-mass stellar models and isochrones with
  self-consistent treatment of rotation. From 0.2 to 1.5
  M$_{{\ensuremath{\odot}}}$ at seven metallicities from PMS to TAMS}}.
\bjtitle{Astronomy and Astrophysics}
\bvolume{631},
\bfpage{77}
(\byear{2019}).
\doiurl{10.1051/0004-6361/201935160}
\end{barticle}
\endbibitem

\bibitem{Jao2018}
\begin{barticle}
\bauthor{\bsnm{Jao}, \binits{W.-C.}},
\bauthor{\bsnm{Henry}, \binits{T.J.}},
\bauthor{\bsnm{Gies}, \binits{D.R.}},
\bauthor{\bsnm{Hambly}, \binits{N.C.}}:
\batitle{{A Gap in the Lower Main Sequence Revealed by Gaia Data Release 2}}.
\bjtitle{Astrophysical Journall}
\bvolume{861}(\bissue{1}),
\bfpage{11}
(\byear{2018}).
\doiurl{10.3847/2041-8213/aacdf6}
\end{barticle}
\endbibitem

\bibitem{gaia}
\begin{barticle}
\bauthor{\bsnm{Collaboration}, \binits{G.}}:
\batitle{{The Gaia mission}}.
\bjtitle{Astronomy and Astrophysics}
\bvolume{595},
\bfpage{1}
(\byear{2016}).
\doiurl{10.1051/0004-6361/201629272}
\end{barticle}
\endbibitem

\bibitem{gaia2018}
\begin{barticle}
\bauthor{\bsnm{Collaboration}, \binits{G.}}:
\batitle{{Gaia Data Release 2. Summary of the contents and survey properties}}.
\bjtitle{Astronomy and Astrophysics}
\bvolume{616},
\bfpage{1}
(\byear{2018}).
\doiurl{10.1051/0004-6361/201833051}
\end{barticle}
\endbibitem

\bibitem{vansanders2012}
\begin{barticle}
\bauthor{\bparticle{van} \bsnm{Saders}, \binits{J.L.}},
\bauthor{\bsnm{Pinsonneault}, \binits{M.H.}}:
\batitle{{An $^{3}$He-driven Instability near the Fully Convective Boundary}}.
\bjtitle{Astrophysical Journal}
\bvolume{751}(\bissue{2}),
\bfpage{98}
(\byear{2012}).
\doiurl{10.1088/0004-637X/751/2/98}
\end{barticle}
\endbibitem

\bibitem{Baraffe2018}
\begin{barticle}
\bauthor{\bsnm{Baraffe}, \binits{I.}},
\bauthor{\bsnm{Chabrier}, \binits{G.}}:
\batitle{{A closer look at the transition between fully convective and partly
  radiative low-mass stars}}.
\bjtitle{Astronomy and Astrophysics}
\bvolume{619},
\bfpage{177}
(\byear{2018}).
\doiurl{10.1051/0004-6361/201834062}
\end{barticle}
\endbibitem

\bibitem{MacDonald2018}
\begin{barticle}
\bauthor{\bsnm{MacDonald}, \binits{J.}},
\bauthor{\bsnm{Gizis}, \binits{J.}}:
\batitle{{An explanation for the gap in the Gaia HRD for M dwarfs}}.
\bjtitle{Monthly Notices of the Royal Astronomical Society}
\bvolume{480}(\bissue{2}),
\bfpage{1711}--\blpage{1714}
(\byear{2018}).
\doiurl{10.1093/mnras/sty1888}
\end{barticle}
\endbibitem

\bibitem{Feiden2021}
\begin{barticle}
\bauthor{\bsnm{Feiden}, \binits{G.A.}},
\bauthor{\bsnm{Skidmore}, \binits{K.}},
\bauthor{\bsnm{Jao}, \binits{W.-C.}}:
\batitle{{Gaia Gaps and the Physics of Low-mass Stars. I. The Fully Convective
  Boundary}}.
\bjtitle{Astrophysical Journal}
\bvolume{907}(\bissue{1}),
\bfpage{53}
(\byear{2021}).
\doiurl{10.3847/1538-4357/abcc03}
\end{barticle}
\endbibitem

\bibitem{Skumanich1972}
\begin{barticle}
\bauthor{\bsnm{{Skumanich}}, \binits{A.}}:
\batitle{{Time Scales for CA II Emission Decay, Rotational Braking, and Lithium
  Depletion}}.
\bjtitle{Astrophysical Journal}
\bvolume{171},
\bfpage{565}
(\byear{1972}).
\doiurl{10.1086/151310}
\end{barticle}
\endbibitem

\bibitem{Curtis2020}
\begin{barticle}
\bauthor{\bsnm{Curtis}, \binits{J.L.}},
\bauthor{\bparticle{et} \bsnm{al.}}:
\batitle{{When Do Stalled Stars Resume Spinning Down? Advancing Gyrochronology
  with Ruprecht 147}}.
\bjtitle{Astrophysical Journal}
\bvolume{904}(\bissue{2}),
\bfpage{140}
(\byear{2020}).
\doiurl{10.3847/1538-4357/abbf58}
\end{barticle}
\endbibitem

\bibitem{Gordon2021}
\begin{barticle}
\bauthor{\bsnm{Gordon}, \binits{T.A.}},
\bauthor{\bparticle{et} \bsnm{al.}}:
\batitle{{Stellar Rotation in the K2 Sample: Evidence for Modified Spin-down}}.
\bjtitle{Astrophysical Journal}
\bvolume{913}(\bissue{1}),
\bfpage{70}
(\byear{2021}).
\doiurl{10.3847/1538-4357/abf63e}
\end{barticle}
\endbibitem

\bibitem{Mestel1984}
\begin{bchapter}
\bauthor{\bsnm{{Mestel}}, \binits{L.}}:
\bctitle{{Angular Momentum Loss During Pre-Main Sequence Contraction}}.
In: \beditor{\bsnm{{Baliunas}}, \binits{S.L.}},
\beditor{\bsnm{{Hartmann}}, \binits{L.}} (eds.)
\bbtitle{Cool Stars, Stellar Systems, and the Sun}
vol. \bseriesno{193},
p. \bfpage{49}
(\byear{1984}).
\doiurl{10.1007/3-540-12907-3_179}
\end{bchapter}
\endbibitem

\bibitem{Wood2021}
\begin{barticle}
\bauthor{\bsnm{Wood}, \binits{B.E.}},
\bauthor{\bparticle{et} \bsnm{al.}}:
\batitle{{New Observational Constraints on the Winds of M dwarf Stars}}.
\bjtitle{Astrophysical Journal}
\bvolume{915}(\bissue{1}),
\bfpage{37}
(\byear{2021}).
\doiurl{10.3847/1538-4357/abfda5}
\end{barticle}
\endbibitem

\bibitem{Anthony2022}
\begin{barticle}
\bauthor{\bsnm{Anthony}, \binits{F.}},
\bauthor{\bparticle{et} \bsnm{al.}}:
\batitle{{Activity and Rotation of Nearby Field M Dwarfs in the TESS Southern
  Continuous Viewing Zone}}.
\bjtitle{Astronomical Journal}
\bvolume{163}(\bissue{6}),
\bfpage{257}
(\byear{2022}).
\doiurl{10.3847/1538-3881/ac6110}
\end{barticle}
\endbibitem

\bibitem{See2019}
\begin{barticle}
\bauthor{\bsnm{{See}}, \binits{V.}},
\bauthor{\bsnm{{Matt}}, \binits{S.P.}},
\bauthor{\bsnm{{Finley}}, \binits{A.J.}},
\bauthor{\bsnm{{Folsom}}, \binits{C.P.}},
\bauthor{\bsnm{{Boro Saikia}}, \binits{S.}},
\bauthor{\bsnm{{Donati}}, \binits{J.-F.}},
\bauthor{\bsnm{{Fares}}, \binits{R.}},
\bauthor{\bsnm{{H{\'e}brard}}, \binits{{\'E}.M.}},
\bauthor{\bsnm{{Jardine}}, \binits{M.M.}},
\bauthor{\bsnm{{Jeffers}}, \binits{S.V.}},
\bauthor{\bsnm{{Marsden}}, \binits{S.C.}},
\bauthor{\bsnm{{Mengel}}, \binits{M.W.}},
\bauthor{\bsnm{{Morin}}, \binits{J.}},
\bauthor{\bsnm{{Petit}}, \binits{P.}},
\bauthor{\bsnm{{Vidotto}}, \binits{A.A.}},
\bauthor{\bsnm{{Waite}}, \binits{I.A.}},
\bauthor{\bsnm{{BCool Collaboration}}}:
\batitle{{Do Non-dipolar Magnetic Fields Contribute to Spin-down Torques?}}
\bjtitle{\apj}
\bvolume{886}(\bissue{2}),
\bfpage{120}
(\byear{2019})
{\href{https://arxiv.org/abs/1910.02129}{{arXiv:1910.02129}}}
{[astro-ph.SR]}.
\doiurl{10.3847/1538-4357/ab46b2}
\end{barticle}
\endbibitem

\bibitem{See2020}
\begin{barticle}
\bauthor{\bsnm{{See}}, \binits{V.}},
\bauthor{\bsnm{{Lehmann}}, \binits{L.}},
\bauthor{\bsnm{{Matt}}, \binits{S.P.}},
\bauthor{\bsnm{{Finley}}, \binits{A.J.}}:
\batitle{{How Much Do Underestimated Field Strengths from Zeeman-Doppler
  Imaging Affect Spin-down Torque Estimates?}}
\bjtitle{\apj}
\bvolume{894}(\bissue{1}),
\bfpage{69}
(\byear{2020})
{\href{https://arxiv.org/abs/2002.11774}{{arXiv:2002.11774}}}
{[astro-ph.SR]}.
\doiurl{10.3847/1538-4357/ab7918}
\end{barticle}
\endbibitem

\bibitem{OSG1}
\begin{bchapter}
\bauthor{\bsnm{Pordes}, \binits{R.}},
\bauthor{\bparticle{et} \bsnm{al.}}:
\bctitle{The open science grid}.
In: \bbtitle{J. Phys. Conf. Ser.}
\bsertitle{78},
vol. \bseriesno{78},
p. \bfpage{012057}
(\byear{2007}).
\doiurl{10.1088/1742-6596/78/1/012057}
\end{bchapter}
\endbibitem

\bibitem{OSG2}
\begin{bchapter}
\bauthor{\bsnm{Sfiligoi}, \binits{I.}},
\bauthor{\bparticle{et} \bsnm{al.}}:
\bctitle{The pilot way to grid resources using glideinwms}.
In: \bbtitle{2009 WRI World Congress on Computer Science and Information
  Engineering}.
\bsertitle{2},
vol. \bseriesno{2},
pp. \bfpage{428}--\blpage{432}
(\byear{2009}).
\doiurl{10.1109/CSIE.2009.950}
\end{bchapter}
\endbibitem

\bibitem{vizier}
\begin{barticle}
\bauthor{\bsnm{Ochsenbein}, \binits{F.}},
\bauthor{\bsnm{Bauer}, \binits{P.}},
\bauthor{\bsnm{Marcout}, \binits{J.}}:
\batitle{{The VizieR database of astronomical catalogues}}.
\bjtitle{Astronomy and Astrophysics, Supplement}
\bvolume{143},
\bfpage{23}--\blpage{32}
(\byear{2000})
{\href{https://arxiv.org/abs/astro-ph/0002122}{{astro-ph/0002122}}}.
\doiurl{10.1051/aas:2000169}
\end{barticle}
\endbibitem

\bibitem{simbad}
\begin{barticle}
\bauthor{\bsnm{Wenger}, \binits{M.}},
\bauthor{\bparticle{et} \bsnm{al.}}:
\batitle{{The SIMBAD astronomical database. The CDS reference database for
  astronomical objects}}.
\bjtitle{Astronomy and Astrophysics, Supplement}
\bvolume{143},
\bfpage{9}--\blpage{22}
(\byear{2000})
{\href{https://arxiv.org/abs/astro-ph/0002110}{{astro-ph/0002110}}}.
\doiurl{10.1051/aas:2000332}
\end{barticle}
\endbibitem

\bibitem{Green2018}
\begin{barticle}
\bauthor{\bsnm{Green}, \binits{G.M.}}:
\batitle{{dustmaps: A Python interface for maps of interstellar dust}}.
\bjtitle{The Journal of Open Source Software}
\bvolume{3}(\bissue{26}),
\bfpage{695}
(\byear{2018}).
\doiurl{10.21105/joss.00695}
\end{barticle}
\endbibitem

\bibitem{Green20182}
\begin{barticle}
\bauthor{\bsnm{Green}, \binits{G.M.}},
\bauthor{\bparticle{et} \bsnm{al.}}:
\batitle{{Galactic reddening in 3D from stellar photometry - an improved map}}.
\bjtitle{Monthly Notices of the Royal Astronomical Society}
\bvolume{478}(\bissue{1}),
\bfpage{651}--\blpage{666}
(\byear{2018}).
\doiurl{10.1093/mnras/sty1008}
\end{barticle}
\endbibitem

\bibitem{Yu2018}
\begin{barticle}
\bauthor{\bsnm{Yu}, \binits{J.}},
\bauthor{\bsnm{Liu}, \binits{C.}}:
\batitle{{The age-velocity dispersion relation of the Galactic discs from
  LAMOST-Gaia data}}.
\bjtitle{Monthly Notices of the Royal Astronomical Society}
\bvolume{475}(\bissue{1}),
\bfpage{1093}--\blpage{1103}
(\byear{2018}).
\doiurl{10.1093/mnras/stx3204}
\end{barticle}
\endbibitem

\bibitem{kepler}
\begin{barticle}
\bauthor{\bsnm{Borucki}, \binits{W.J.}},
\bauthor{\bparticle{et} \bsnm{al.}}:
\batitle{{Kepler Planet-Detection Mission: Introduction and First Results}}.
\bjtitle{Science}
\bvolume{327}(\bissue{5968}),
\bfpage{977}
(\byear{2010}).
\doiurl{10.1126/science.1185402}
\end{barticle}
\endbibitem

\bibitem{ztf}
\begin{barticle}
\bauthor{\bsnm{Bellm}, \binits{E.C.}},
\bauthor{\bparticle{et} \bsnm{al.}}:
\batitle{{The Zwicky Transient Facility: System Overview, Performance, and
  First Results}}.
\bjtitle{Publications of the ASP}
\bvolume{131}(\bissue{995}),
\bfpage{018002}
(\byear{2019}).
\doiurl{10.1088/1538-3873/aaecbe}
\end{barticle}
\endbibitem

\bibitem{Gaia2022}
\begin{botherref}
\oauthor{\bsnm{Collaboration}, \binits{G.}}:
{Gaia Data Release 3: Summary of the content and survey properties}.
arXiv e-prints,
2208--00211
(2022)
{\href{https://arxiv.org/abs/2208.00211}{{arXiv:2208.00211}}}
{[astro-ph.GA]}
\end{botherref}
\endbibitem

\bibitem{astropy:2013}
\begin{barticle}
\bauthor{\bsnm{Collaboration}, \binits{A.}}:
\batitle{{Astropy: A community Python package for astronomy}}.
\bjtitle{Astronomy and Astrophysics}
\bvolume{558},
\bfpage{33}
(\byear{2013}).
\doiurl{10.1051/0004-6361/201322068}
\end{barticle}
\endbibitem

\bibitem{astropy:2018}
\begin{barticle}
\bauthor{\bsnm{Price-Whelan}, \binits{A.M.}},
\bauthor{\bparticle{et} \bsnm{al.}}:
\batitle{{The Astropy Project: Building an Open-science Project and Status of
  the v2.0 Core Package}}.
\bjtitle{Astronomical Journal}
\bvolume{156},
\bfpage{123}
(\byear{2018}).
\doiurl{10.3847/1538-3881/aabc4f}
\end{barticle}
\endbibitem

\bibitem{Asplund2021}
\begin{barticle}
\bauthor{\bsnm{Asplund}, \binits{M.}},
\bauthor{\bsnm{Amarsi}, \binits{A.M.}},
\bauthor{\bsnm{Grevesse}, \binits{N.}}:
\batitle{{The chemical make-up of the Sun: A 2020 vision}}.
\bjtitle{Astronomy and Astrophysics}
\bvolume{653},
\bfpage{141}
(\byear{2021}).
\doiurl{10.1051/0004-6361/202140445}
\end{barticle}
\endbibitem

\bibitem{KS66}
\begin{barticle}
\bauthor{\bsnm{Krishna~Swamy}, \binits{K.S.}}:
\batitle{{Profiles of Strong Lines in K-Dwarfs}}.
\bjtitle{Astrophysical Journal}
\bvolume{145},
\bfpage{174}
(\byear{1966}).
\doiurl{10.1086/148752}
\end{barticle}
\endbibitem

\bibitem{Matt2012}
\begin{barticle}
\bauthor{\bsnm{{Matt}}, \binits{S.P.}},
\bauthor{\bsnm{{MacGregor}}, \binits{K.B.}},
\bauthor{\bsnm{{Pinsonneault}}, \binits{M.H.}},
\bauthor{\bsnm{{Greene}}, \binits{T.P.}}:
\batitle{{Magnetic Braking Formulation for Sun-like Stars: Dependence on Dipole
  Field Strength and Rotation Rate}}.
\bjtitle{\apjl}
\bvolume{754}(\bissue{2}),
\bfpage{26}
(\byear{2012})
\end{barticle}
\endbibitem

\bibitem{Cranmer2011}
\begin{barticle}
\bauthor{\bsnm{{Cranmer}}, \binits{S.R.}},
\bauthor{\bsnm{{Saar}}, \binits{S.H.}}:
\batitle{{Testing a Predictive Theoretical Model for the Mass Loss Rates of
  Cool Stars}}.
\bjtitle{\apj}
\bvolume{741}(\bissue{1}),
\bfpage{54}
(\byear{2011})
\end{barticle}
\endbibitem

\end{thebibliography}


\begin{appendices}

\section{Method}\label{secA1}
\subsection{Gyro-kinematic Age Sample}\label{subsection:method}
We determine gyro-kinematic ages following the procedure described in Lu et al. 2021 \cite{Lu2021}, where the vertical velocity dispersion for each star is calculated from vertical velocities of stars that are similar in temperature (\teff; calculated from $G_{\rm BP}-G_{\rm RP}$ measurements using a polynomial fit taken from Curtis et al. 2020 \cite{Curtis2020}; $G_{\rm BP}-G_{\rm RP}$ de-reddened using {\tt dustmap} \citep{Green2018, Green20182}, rotation periods ($P_{\rm rot}$), absolute Gaia magnitude ($M_G$; extinction-corrected using {\tt dustmap}), and Rossby number ($Ro$; See et al. in prep.) to the targeted star. 
We then converted the velocity dispersions into stellar ages using an age-velocity-dispersion relation in \cite{Yu2018}.
Post-main-sequence stars are cut out by only selecting stars with $M_G > 4.2$ mag, and equal-mass binaries are excluded by fitting a 6\ith-order polynomial to the main-sequence stars in $M_G$-$T_{\rm eff}$ space, moving the fitted polynomial along the $M_G$-axis so that it lies right below the equal-mass binary sequence, and removing stars with $M_G$ greater than the modified polynomial. 

The dataset used in this work is from Lu et al. (in prep.) that combines the $\sim$ 20,000 stars in Lu et al. 2021 \cite{Lu2021} from Kepler \citep{kepler} and the $\sim$ 50,000 stars with period measurements from Lu et al. 2022 \cite{Lu2022} and Lu et al. (in prep) from ZTF \citep{ztf}.
The vertical velocities for the ZTF stars are obtained using radial velocity measurements from Gaia DR3 \citep{Gaia2022}. We do this by transforming from the Solar system barycentric ICRS reference frame to Galactocentric Cartesian and cylindrical coordinates using {\tt astropy} \citep{astropy:2013, astropy:2018}.
The bin size is [$T_{\rm eff}$, log$_{10}(P_{\rm rot})$, $Ro$, $M_G$] = [177.8 K, 0.15, 0.15, 0.2 mag], optimized followed the methodology as the one outlined in \cite{Lu2021} using clusters stars ranging from 0.6 - 4 Gyr \citep{Curtis2020, Dungee2022}. 

\subsection{Stellar evolution model}\label{subsection:model}
The stellar evolution model for this work is computed using STAREVOL \citep{Siess2000,Amard2019}.
We used a refined grid of standard models for masses between 0.3 and 0.4 solar masses by steps of 0.01 for 8 metallicities between [Fe/H]=-1 and +0.3. 
Abundances are taken from Asplund et al. 2021 \cite{Asplund2021}, we use an analytical surface atmospheric fit \citep{KS66}, and a solar-calibrated mixing length parameter $\alpha=2.11$.

\subsection{Eliminating Possible Systematic Causes}
To eliminate the possibility that systematics or biases artificially produced this result, we visually examined 100 random ZTF light curves for the fully convective stars between 3300 K to 3500 K (where the double spin-down sequence exists) with measured periods $>$ 50 days.
We found no systematic in the measured periods that could cause all the fully convective stars to be on a period harmonic. 
Moreover, rotation periods in the spin-down sequence for the fully convective stars are not period harmonics (integer multiples) of those for the partially convective stars (Fig.\ref{fig:1}). 
It is also unlikely that a bias or systematic in rotation period measurements would conspire to create a double-sequence only  around the fully convective boundary (Fig.\ref{fig:2}) predicted by the stellar evolution model (Fig.\ref{fig:1}), and that the CMD gap is able to nicely separate the two sequences (Fig.\ref{fig:2} and Fig.\ref{fig:3}). 

To eliminate the possibility of a systematic effect that could caused by combining the two data sets from Kepler and ZTF, we performed the same tests with just the ZTF sample with RV measurements from Gaia DR3 and found no significant changes in the results presented in this paper.

\subsection{Limitations of the gyro-kinematic age-dating method}
The gyro-kinematic age dating technique assumes that stars with similar parameters (effective temperature, rotation period, Rossby number, absolute magnitude) are approximately the same age. In general, it is expected that this assumption should hold. However, if partially convective and fully convective star2s have different braking laws, this assumption may be broken for stars near the fully convective boundary. Stars on either side of the boundary, with otherwise similar properties, could have quite different ages. It is therefore worth examining the behaviour of the gyro-kinematic age dating method at the fully convective boundary in more detail given that our results rest upon this technique.

Let us consider a star near the fully convective boundary; whether it is on the fully or partially convective side of the boundary does not matter. To estimate a gyro-kinematic age for this star, we would calculate the vertical velocity dispersion of all the stars contained in a bin in parameter space centered on the star in question and then convert this to an age estimate using an age-velocity-dispersion relation. Since we are near the fully convective boundary, this bin could contain both fully convective and partially convective stars. If fully convective and partially convective stars do indeed obey different braking laws, then the fully convective and partially convective stars likely have different ages even though they are contained within the same bin. The estimated age for the star will therefore be an averaged age of the fully convective and partially convective populations. This would have the effect of smearing out the sequences in Fig.\ref{fig:1} (left) and Fig.\ref{fig:2}. That we still see two sequences in these figures, even after this smearing out, suggests that our results are robust to this limitation of the gyro-kinematic age dating technique. Indeed, in reality, the sequences could be even more well-defined than shown in this work.

\subsection{Torques for FC and PC stars spinning at the same rate}\label{subsection:calculation}
In this section, we determine how much larger the braking torque acting on FC stars is compared to PC stars by considering their observed rotation evolution. At late ages, it is thought that stellar rotation periods increase with the square root of age, i.e. the well known Skumanich relation,

\begin{equation}
    P(t) = \alpha t^{0.5},
    \label{eq:Skumanich}
\end{equation}
or in terms of angular frequency,

\begin{equation}
    \Omega (t) = \frac{2\pi}{\alpha} t^{-0.5}.
    \label{eq:OmegaSkumanich}
\end{equation}
Here, $\alpha$ is a constant of proportionality that we will empirically determine later. The rotation evolution of low-mass stars is governed by the angular momentum equation, 

\begin{equation}
        \frac{d\Omega}{dt} = \frac{T}{I} - \frac{\Omega}{I}\frac{dI}{dt},
    \label{Eq:AngMomEq}
\end{equation}
where $T$ is the spin-down torque and $I$ is the moment of inertia. On the main sequence, the $dI/dt$ term is approximately zero since a star's stellar structure does not appreciable change during this phase of evolution. Therefore, differentiating eq. (\ref{eq:OmegaSkumanich}), substituting into eq. (\ref{Eq:AngMomEq}) and rearranging, one finds that the spin-down torque is given by 

\begin{equation}
    T = -\frac{\alpha^2 I}{8\pi^2} \Omega^3.
\end{equation}
The ratio of the torques acting on FC and PC stars, assuming the same angular frequency, is therefore

\begin{equation}
    \frac{T_{\rm FC}}{T_{\rm PC}} = \left(\frac{\alpha_{\rm FC}}{\alpha_{\rm PC}}\right)^2 \frac{I_{\rm FC}}{I_{\rm PC}}.
\end{equation}
From stars in the range 3400K $<T_{\rm eff}<$ 3500K (Fig. \ref{fig:2}), we can see that $\alpha_{\rm FC}$/$\alpha_{\rm PC}$ is $\sim$ 1.5. 
As a result, if we consider the case where the FC and PC stars are both almost exactly on the fully convective boundary, then $I_{\rm FC}/I_{\rm PC} \sim 1$ and we find that the ratio of torques is $T_{\rm FC}/T_{\rm PC} \sim 1.5^2 = 2.25$.

The stellar-wind torque formulation \cite{Matt2012} predicts that (for fixed stellar mass, radius, and rotation rate) a factor of 2.25 larger torque requires a dipole field strength that is larger by a factor of $\sim$2.5, a mass loss rate that is $\sim$4.2 times larger, or some combination of both of those factors.  The magnetic activity that is responsible for driving mass loss rates is likely dependent on small-scale magnetic structures on the stellar surface \cite{Cranmer2011}. 
Thus an enhanced stellar wind torque could indicate an enhancement of either or both large-scale and small-scale fields.

{\bf Data availability}. 
The rest of the relevant datasets are available from the corresponding author upon reasonable request.

{\bf Code availability}. No new codes are developed in this paper.



\end{appendices}


\end{document}